\documentclass[11pt]{article}
\usepackage{blois,epsfig}
\usepackage[latin1]{inputenc}
\usepackage[OT1]{fontenc}

\def\be{\begin{equation}}
\def\ee{\end{equation}}
\def\bea{\begin{eqnarray}}
\def\eea{\end{eqnarray}}

\begin{document}
\vspace*{4cm}
\title{Ultra High Energy Cosmic Ray Puzzle and the \\
Plasma Wakefield Acceleration }

\author{Feng-Yin Chang$^{1,2}$, Pisin Chen$^{1,3,4}$, Guey-Lin Lin$^{2,3}$, Robert J. Noble$^5$, Kevin Reil$^{1}$, and Richard Sydora$^{6}$ }

\address{$^1$Kavli Institute for Particle Astrophysics and
Cosmology, Stanford Linear Accelerator Center, Stanford
University, Stanford, CA 94305, USA.}
\address{$^2$Institute of
Physics, National Chiao-Tung University, Hsinchu 300, Taiwan.}
\address{$^3$Leung Center
for Cosmology and Particle Astrophysics, National Taiwan University,
Taipei 106, Taiwan.}
\address{$^4$Institute for Astrophysics, National Taiwan
University, Taipei 106, Taiwan.}
\address{$^5$Stanford Linear Accelerator Center, Stanford
University, Stanford, CA 94305, USA.}
\address{$^6$Department of Physics, University of Alberta,
Edmonton, Alberta, Canada.}
\maketitle\abstracts{Magnetowave induced plasma wakefield acceleration (MPWA) in a
relativistic astrophysical outflow has been proposed as a viable
mechanism for the acceleration of cosmic particles to ultra high
energies. Here we present simulation results
that demonstrate the viability of this mechanism. We invoke the high frequency and high speed whistler
mode for the driving pulse. The plasma wakefield so induced
validates precisely the theoretical prediction. This mechanism is shown capable of accelerating charged particles to ZeV energies in Active Galactic Nuclei (AGN).
}

\section{Introduction}

The origin of ultra high energy cosmic rays (UHECR) is an intriguing question in astrophysics. Theories categorized as either ``top-down" or ``bottom-up" scenarios are proposed to answer this question.
Each scenario faces its own theoretical and observational challenges \cite{Olinto}.
Since the observations from HiRes \cite{cosmic:HiRes} and
Auger \cite{cosmic:Auger} confirm the Greisen-Zatepin-Kuz'min (GZK) suppression of the cosmic ray flux \cite{gzk}, the need for top-down exotic models is reduced. Hence the challenge to find a viable ``bottom up" mechanism for accelerating UHECR becomes more acute.

Shocks, unipolar inductors and magnetic flares are the three most
potent, observed, ``conventional" accelerators that can be
extended to account for $\sim{\rm ZeV} (=10^{21}{\rm eV})$ energy
cosmic rays \cite{Blandford:1999}. Radio jet termination shocks and
gamma ray bursts (GRB) have been invoked as sites for the shock
acceleration, while dormant galactic center black holes and
magnetars have been proposed as sites for the unipolar inductor
acceleration and the flare acceleration, respectively. Each of
these models, however, presents problems \cite{Blandford:1999}.
Evidently, novel acceleration mechanisms that can avoid the
difficulties faced by these conventional models should not be
overlooked.

Plasma wakefield accelerators \cite{LWFA:Tajima,PWFA:Chen85} are
known to possess two salient features: (1) The energy gain per
unit distance does not depend (inversely) on the particle's
instantaneous energy;
(2) The acceleration is linear. These features are essential for a good acceleration efficiency.
Although high-intensity, ultra-short photon or particle beam
pulses that excite the laboratory plasma wakefields are not available in the astrophysical setting, large amplitude plasma
wakefields can instead be excited by the astrophysically
abundant plasma ``magnetowaves" \cite{plasma:Chen02}. Protons can be accelerated beyond
ZeV energy by riding on such wakefields. This attractive
concept has never been validated through self-consistent computer
simulations. In this presentation, we report our simulation that confirm this concept \cite{Chang:2007um}. We also discuss this acceleration mechanism in AGN.

\section{Wakefield Excited by the Magnetowave: Theory and Simulation}

To ensure the linear acceleration, we consider wave modes propagating parallel to the external magnetic
field. In this case, the eigenmodes are circularly polarized with
dispersion relations given by \cite{Stix}
\begin{equation}
 \omega^{2}=k^{2}c^{2}+\frac{\omega_{ip}^{2}}{1\pm\omega_{ic}/\omega}+
\frac{\omega_{p}^{2}}{1\mp\omega_{c}/\omega}\, ,\label{dispersion}
\end{equation}
where the upper (lower) signs denote the right-hand (left-hand)
circularly polarized waves, the subscript $i$ denotes the ion
species, $\omega_{p}$ and $\omega_{c}$ are the electron plasma frequency and the
electron cyclotron frequency respectively.
The right-hand
polarized, low frequency solution is called the whistler wave
which propagates at a phase velocity less than the speed of light.
For a sufficiently strong magnetic field such that
$\omega_{c}\gg \omega_{p}$, the dispersion of the whistler mode
becomes more linear over a wider range of wavenumbers with phase
velocity approaching the speed of light (see Fig.1).  In this
regime the traveling wave pulses can maintain their shape over
macroscopic distance, a condition desirable for plasma wakefield
acceleration.
\begin{figure}[htb]
\begin{center}
\psfig{figure=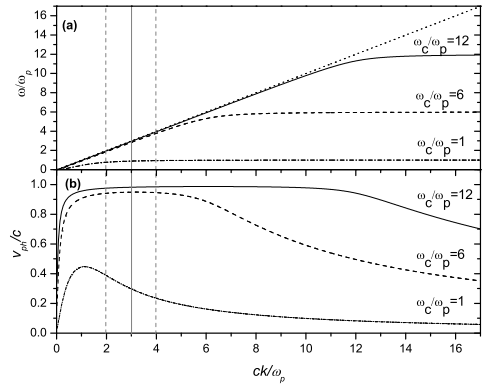,height=2.0in}
\caption{(a) Frequency and (b) phase velocity versus wavenumber
for different magnetic fields. The vertical solid line is
the mean value of the pulse wavenumber chosen for the PIC
simulation, and the dashed lines its range.
\label{fig:dispersion}}
\end{center}
\end{figure}

The whistler wave packet induces a ponderomotive force along the external magnetic field direction \cite{wk}. For convenience, this direction is taken to be along the $+z$ axis. The ponderomotive force leads to the plasma wakefield $E_z$ which, in the co-moving coordinate $\zeta \equiv z-v_gt$, reads:
\begin{eqnarray}
E_z(\zeta)=-\frac{ek_pE_{_W}^2}
{m_e\omega(\omega-\omega_c)}\left[1+\frac{kv_g\omega_c}{\omega(\omega-\omega_c)}\right]
\chi(\zeta), \label{wakefield}
\end{eqnarray}
where $k_p=\omega_p/v_g$ and the form factor $\chi(\zeta)$ is given by
\begin{equation}
\chi(\zeta)=\frac{k_p}{2E_{_W}^2}\int_{\zeta}^{\infty}d\zeta^{\prime}E_{_W}^2(\zeta^{\prime})
\cos\left[k_p(\zeta-\zeta^{\prime})\right],
\end{equation}
with $E_{_W}(\zeta)$ the field strength of the whistler wave packet and $E_W$ its maximum value.
Assuming the driving pulse
frequency is centered around $\omega$ and its group velocity $v_{g}={\mbox d}\omega/{\mbox d}k\approx
\omega/k$, the maximum wakefield attainable behind the driving pulse is found to be
\begin{equation}
E_z^{\rm max}=\chi\frac{k^{2}c^2}{(\omega-\omega_{c})^{2}} a_0^{2}E_{wb}, \label{gradient}
\end{equation}
for a
Gaussian driving pulse $E_{_W}(\zeta)=E_{_W}
\exp(-\zeta^2/2\sigma^2)$ with $\chi=\sqrt{\pi}k_p\sigma \exp(-k_p^2\sigma^2/4)/2$, $a_0\equiv eE_{_W}/m_ec\omega$ the ``strength parameter" of the driving pulse, and $E_{wb}\equiv m_ec\omega_p/e$  the ``wavebreaking" field.
We note that Eq.~\ref{wakefield} was derived under the non-relativistic approximation for the electron motion in the plasma. The relativistic generalization of this equation deserves further studies.

We have conducted computer simulations to study the MPWA process
driven by a Gaussian driving whistler pulse described above. Our
simulation model integrates the relativistic Newton-Lorentz
equations of motion in the self-consistent electric and magnetic
fields determined by the solution to Maxwell's
equations \cite{Dawson,Sydora}.
We used
a wavepacket with Gaussian width $\sigma =80\Delta/\sqrt{2}$,
where $\Delta$ is the cell size taken to be unity, and the
wavenumber $k = 2\pi/60\Delta$. The physical parameters
$\omega_{c}/ \omega_{p}=6$, $m_i/m_e=2000$ and electron collisionless skin depth
$c/\omega_{p}=30\Delta$ were taken in the simulations. Other numerical parameters used
are: total number of cells in the $z$-direction, $L_z
=8192\Delta=273c/\omega_p$, average number of particles per cell
was 10, and the time step $\omega_p\Delta t=0.1$. The fields were
normalized by $(1/30)E_{wb}$.

We set the maximum amplitude $E_{_W}=10$, which gives the
strength parameter $a_{0}=eE_{_W}/m_ec\omega=0.11$. The
pulse was initialized at $z_0=500\Delta=16.66 c/\omega_p$. To
avoid spurious effects, we gradually ramped up the driving pulse
amplitude until $t=100 \omega_{p}^{-1}$, during which the plasma
feedback to the driving pulse was ignored. After this time, the
driving pulse-plasma interaction was tracked self-consistently. As
the dispersion relation in this regime is not perfectly linear,
there was a gradual spread of the pulse width. Thus $\chi$ and
$E_{_W}$ of the driving pulse decrease accordingly. As a result,
the maximum wakefield amplitude, $E_z^{\rm max}$, declined in time. Even so,
it agrees very well with the theoretical value of $E_{z}^{\rm max}\sim
0.266(1/30)E_{wb}$. Fig.2 is a snapshot of $E_{x}$ and $E_{z}$ at
$t=230\omega_{p}^{-1}$. We note that while the driving pulse
continues to disperse, the wakefield remains extremely coherent.
\begin{figure}[htb]
\begin{center}
\psfig{figure=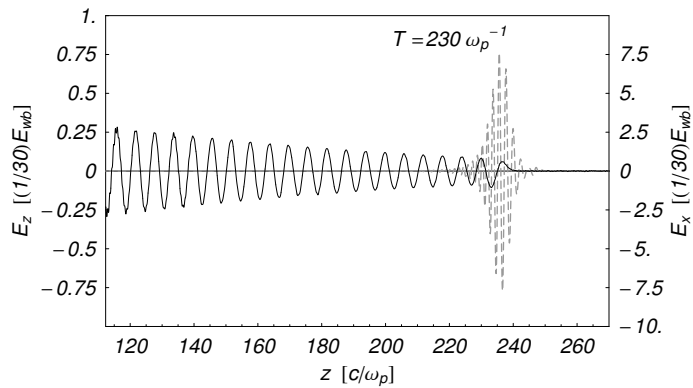,height=2.0in}
\caption{A snapshot of the plasma wakefield, $E_z$ (in black), induced by the
whistler pulse, $E_x$ (in gray).
\label{fig:wakefield}}
\end{center}
\end{figure}

\section{MPWA Production of UHECR in AGN}

The MPWA production of UHECR can be effective
in AGN. Typically, the jet from an
AGN extends a very long distance with negligible
diverging angle. For an AGN with the central black hole mass $\sim 10^8 M_{\odot}$, it is reasonable to assume $n_{\rm AGN}\sim 10^{10}{\rm cm}^{-3}$ and $B_{\rm AGN}\sim 10^4$ G in the core \cite{rees:1981}. If
we further assume that the AGN luminosity approaches
the Eddington limit ($\sim 10^{46}$ erg/s) and that the AGN
jet size is comparable to the Schwarzschild radius of the
central black hole, then we find that $a_0\sim \sqrt{10\eta}$ and
$E_{wb}\sim 10^5$ V/cm with $\eta$ the fraction of total energy
imparted into the magnetowave modes. Since the frequency
of magnetowave in this case lies in the radio wave
region, we assume that the observed AGN radio wave
luminosity \cite{Arshakian:2006pj} is the result of total mode-conversion from
the magnetoshocks at the same frequency. This then
gives $\eta \sim (10^{-3}-10^{-4})$ and consequently $E_z^{\rm max}\sim \mathcal{O}(10^2)$ eV/cm from Eq.~\ref{gradient}. Hence
the acceleration distance to achieve $E \sim \mathcal{O}(10^{21})$ eV is
about 10 pc, which is a tiny fraction of the typical AGN
jet length.

\section{Summary}

Through PIC simulations, we have confirmed the concept of plasma wakefield
excited by a magnetowave in the magnetized plasma. We have demonstrated how such a wakefield may accelerate particles to ZeV energies in AGN.
As a first step, we investigated MPWA in the parallel-field
configuration. Since both poloidal and toroidal field
components are inevitable in astro-jets, we will further
investigate plasma wakefield excitation and acceleration
under the cross-field configuration. Besides, we have limited our discussions in the linear regime $a_0\ll 1$, which is applicable to AGN. However, in other astrophysical settings, such as GRB, the magnetowave could be much stronger such that $a_0\gg 1$. We will investigate plasma wakefield generations in this regime and therefore explore the MWPA production of UHECR in other powerful astrophysical sites.
\section*{Acknowledgments}
GLL appreciates the travel supports from both NCTU Office of Research and Development and
NTU Leung Center
for Cosmology and Particle Astrophysics.

\end{document}